\documentclass[preprint]{aastex}

\shorttitle{Foreground removal by a neural network}
\shortauthors{N{\o}rgaard-Nielsen}
\begin{document}
\title{Foreground removal from \emph{WMAP 7 yr} polarization maps using an MLP neural network}
\author{H. U. N{\o}rgaard-Nielsen}
\affil{National Space Institute (DTU Space),
    Technical University of Denmark,
    Juliane Mariesvej 30, DK-2100 Copenhagen, Denmark}

 \begin{abstract}

 One of the fundamental problems in extracting the cosmic microwave background signal (CMB) from millimeter/submillimeter observations is the pollution by emission from the Milky Way: synchrotron, free-free, and thermal dust emission. To extract the fundamental cosmological parameters from CMB signal, it is mandatory to minimize this pollution since it will create systematic errors in the CMB power spectra. In previous investigations, it has been demonstrated that the neural network method provide high quality CMB maps from temperature data. Here the analysis is extended to polarization maps. As a concrete example, the WMAP 7-year polarization data, the most reliable determination of the polarization properties of the CMB, has been analysed. The analysis has adopted the frequency maps, noise models, window functions and the foreground models as provided by the WMAP Team, and no auxiliary data is included. Within this framework it is demonstrated that the network can extract the CMB polarization signal with no sign of pollution by the polarized foregrounds. The errors in the derived polarization power spectra are improved compared to the errors derived by the WMAP Team.

\end{abstract}

\keywords{Cosmology: cosmic background radiation; Methods: data analysis}
%

%________________________________________________________________

\section{Introduction}

It is well established that the temperature anisotropies in the cosmic microwave background are a powerful tool to study the early phases of the evolution of the Universe. In addition, polarization measurements provide a new window into the physical conditions in that era. The polarization at large angular scales has the potential to provide information about the Universe when it was only $10^{-35}$ s old, and in addition, information about the ionization history of the Universe.

The CMB polarization probes the evolution of the decoupling and reionization phases. Rees\cite{Rees68} predicted the polarization signal shortly after the discovery of the CMB by Penzias and Wilson \cite{penz65}. Since then there have been considerable effort, both theoretical and observational to study this component. An excellent review can be found in Hu and Dodelson \cite{hudo02}.

Polarization measurements are normally given by the Stokes parameters Q and U, since they have straightforward noise properties. Since their definition depends on the chosen coordinate system, they are not well suited for quantifying the polarization anisotropies. In consequence, Q and U are transformed into E and B modes (E for the curl-free and B for the divergence-free components of the polarization field). The E and B mode formalism was introduced by Seljak \cite{selj97}, Kamionkowski et al.\cite{kami97} and Zaldarriage and Seljak \cite{zald97}.

Fundamental symmetries in the production and growth of the polarization signal constrain the possible configurations of the CMB polarization. Scalar (density) perturbations give rise to T(emperature) and E modes, while tensor (gravitational wave) perturbations give rise to T, E and B modes. Both kinds of perturbations can produce polarization patterns in both the decoupling and reionization periods.

If the primordial inhomogeneities were Gaussian in nature, it follows (assuming linear theory) that CMB fluctuations are also Gaussian and fully described by the 4 cross power spectra TT, EE, BB and TE, while the TB and EB power spectra, from parity considerations, vanish (e.g Kamionkowski et al. \cite{kami97})
As emphasized by Hu and Dodelson \cite{hudo02} among others, density perturbations do not produce B modes to first order, therefore a detection of substantial B polarization will be momentous and push us qualitatively forward into new areas of physics.

The Planck mission was successfully launched on May 14, 2009, and all systems have ever since worked according to expectations. An important part of the preparation for the mission was an evaluation of the available algorithms for removing the Galactic foreground signals, based on detailed simulations, called the \emph{Planck Sky Model}(PSM). Comparison of eight investigated methods for extracting the temperature maps can be found in Leach et al. \cite{leac08}

The overall feasibility of using neural networks to extract the CMB signal from temperature millimeter/submillimeter data was demonstrated by N{\o}rgaard-Nielsen and J{\o}rgensen \cite{hunn08}. N{\o}rgaard-Nielsen and Hebert \cite{hunn09} (hereafter NNH) has shown that a simple neural network can improve the foreground removal significantly, applied on exactly the same data as used in the Leach et al.\cite{leac08} investigation. By analysis of the WMAP 5yr temperature data, N{\o}rgaard-Nielsen \cite{hunn10} (hereafter NN10) has shown that a neural network can provide a significantly improved CMB map of about 75 per cent of the sky, without introducing any auxiliary data.

The WMAP polarization data has been analysed in detail by the WMAP team (Gold et al. \cite{gold11}, Larson et al. \cite{lars11}, Komatsu et al. \cite{koma11}, Dunkley et al. \cite{dunk09}, Kogut et al. \cite{kogu07}, Page et al. \cite{page07}, Kogut et al. \cite{kogu03}). In  this paper we will concentrate on demonstrating the capability of neural networks for removing the Galactic foregrounds from the 7yr CMB Q and U maps and the accuracy of the derived power spectra EE, BB, TE, TB, EB, compared mainly to the results obtained by the WMAP Team on the same data. In the modeling of the polarized Galactic foregrounds, only the models discussed by the WMAP Team have been discussed. Since the feasibility of the NN method is the key issue for this paper, the following issues will not be discussed:
possible residual systematic errors in the WMAP data, the physical interpretation of the derived power spectra, the capability of other neural networks than the adopted simple multilayer perceptron.

\section{The WMAP data}

The NASA WMAP mission has scanned the sky for more than 9 years in the following bands: K(23GHz), Ka(33GHz), Q(41GHz), V(61GHHz) and W(94GHz). Detailed description of the mission can be found at http://map.gsfc.nasa.gov/. In this paper the data for the first 7 years have been taken from the following WMAP web-site:

\emph{lambda.gsfc.nasa.gov/product/map/current/m\_products.cfm}.

\subsection{The WMAP temperature maps by the WMAP Team}
The details of the WMAP data reduction have been intensively discussed in the series of papers released simultaneously with the 1yr, 3yr, 5yr and 7yr data releases listed at the above WMAP web-site. Here we will only give a brief summary.

For the temperature data the WMAP Team has developed a simple method for extracting the CMB signal, the so called ´Internal Linear Combination´ (ILC) method. It is simply a linear combination of the 5 frequency maps, with the coefficients determined by minimizing the total variance of the output map, in 12 predefined areas on the sky.

A basic problem for the ILC method is that it does not take into account the known variations in spectral index and differences in the relative contributions of the Galactic foreground components. Due to the statistical properties of this kind of maps, the WMAP Team recommends that they are not used for cosmological investigations.

To derive the CMB TT power spectrum, the WMAP Team removed templates for the Galactic foregrounds from the frequency maps. The difference between the K and Ka maps --- expressed in thermodynamic temperatures --- is in principle free of the CMB signal and used as a template of the synchrotron emission. The free-free emission is estimated from the full sky H$\alpha$ map by Finkbeiner et al. \cite{fink03}, with corrections for the dust extinction by Bennett et al. \cite{benn03}. For the dust emission ´Model 8´ of Finkbeiner et al. \cite{fink02} has been used. The KQ75 mask (covering $\sim$ 75 per cent of sky) has been applied.
The KQ75 mask, used by the WMAP Team and in this investigation, excludes the central part of the Milky Way and bright point source, altogether excluded 25 percent of the sky.

The TT power spectrum has been derived from the combination of the VV, VW and WW cross power spectra.

Many different methods have been applied to extract the CMB temperature signal from the various versions of the WMAP data. It is outside the scope of this paper to give a detailed description of these methods. A comprehensive review has been given by Delabrouille et al. \cite{dela08}.

\subsection{The analysis of the polarization data by the WMAP Team}

Page et al.\cite{page07} present in detail the reduction scheme for the WMAP polarization data. The CMB Q and U data are derived outside the p06 mask (covering the Milky Way including  the North Polar Spur) using the K band Q and U maps as templates for the synchrotron emission. For the dust emission, the FDS model 8 (Finkbeiner et al.\cite{fink02}), combined with polarization directions from stellar observations, has been used.

For \emph{l }$<$ 23, the cosmological model likelihood of the cross power TE spectrum was estimated directly from a template-cleaned V + W band map (temperature) and a template-cleaned  Ka + Q + V band map (polarization). For higher multipole moments, the MASTER quadratic estimator (Hivon et al. \cite{hivo02})was used. The derived power spectrum fits well to the expectations of the optimal $\lambda$CDM model obtained from the TT power spectrum, see Fig. \ref{fig_pow_te}. The traditional definition, \emph{l} * (\emph{l} + 1) $C_{l}$ /2$\pi$, has been applied for all power spectra: TT, EE, BB, TE, TB, EB.

For the 7yr data, Larson et al.\cite{lars11} found for \emph{l} = 2-7:

$\emph{l}(\emph{l}+1)C^{EE}_{l}/2\pi$ = $0.074^{+0.034}_{-0.025}$ $(\mu K)^{2}$ and

$\emph{l}(\emph{l}+1)C^{BB}_{l}/2\pi$ $<$ 0.055  $(\mu K)^{2}$.

The TB (shown in Fig. \ref{fig_pow_tb}) and EB power spectra are consistent with zero, as was found in the previous WMAP data set.

\begin{figure}[!tbp]
\centering
\includegraphics[width=3.5in]{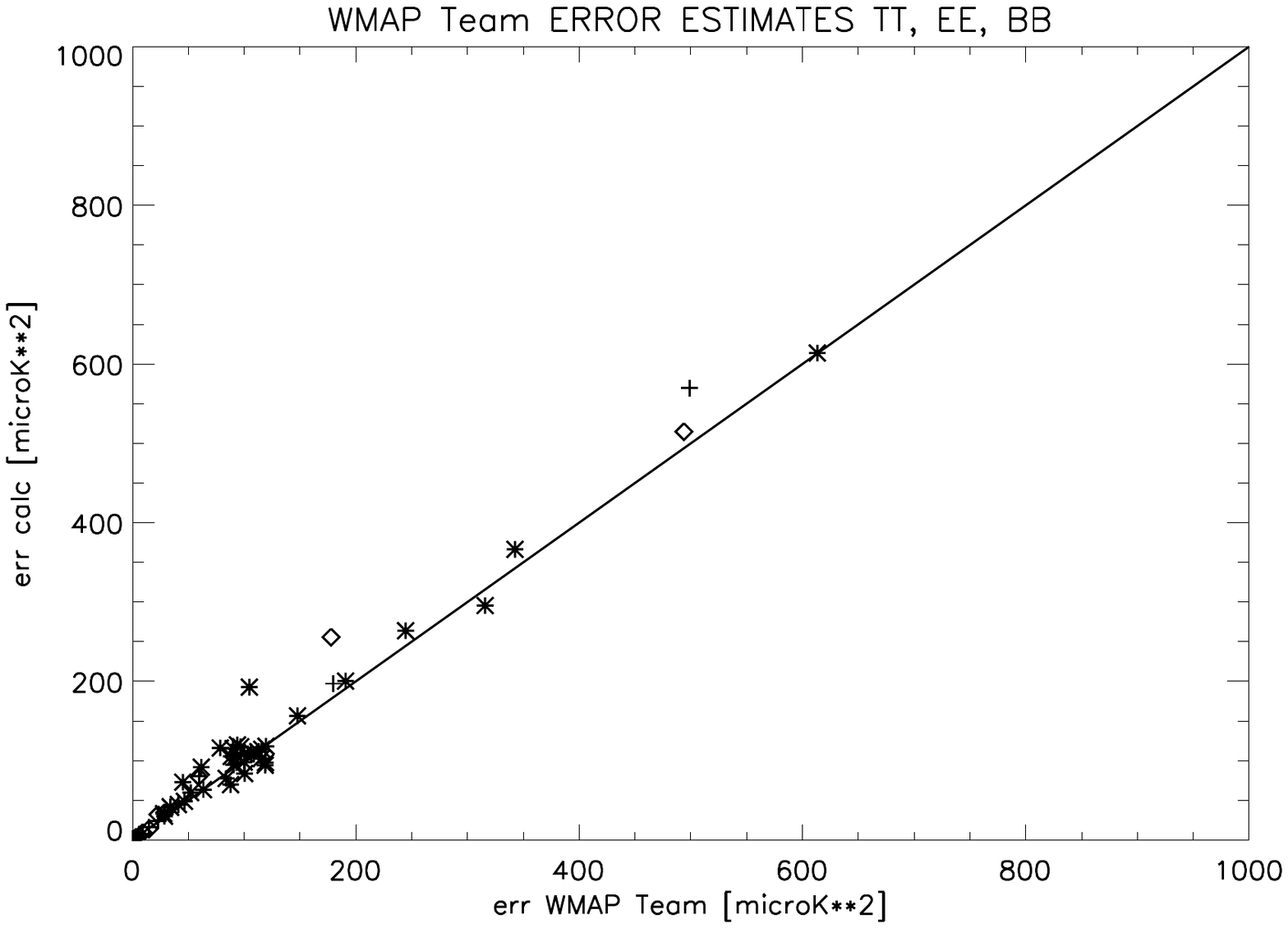}
\caption{The X-axis gives the errors of the power spectra derived by the WMAP Team from a detailed analysis of the noise distributions (incl. observational errors and cosmic variance). The Y-axis gives the errors determined from the scatter within each \emph{l} - interval(more than 5 elements) of the individual power spectra derived by the WMAP Team. The one to one relation is shown. Symbols: TT (asterisks), EE (diamonds).}
\label{fig_sigma_team_tt_ee_bb}
\end{figure}

\begin{figure}[!tbp]
\centering
\includegraphics[width=3.5 in]{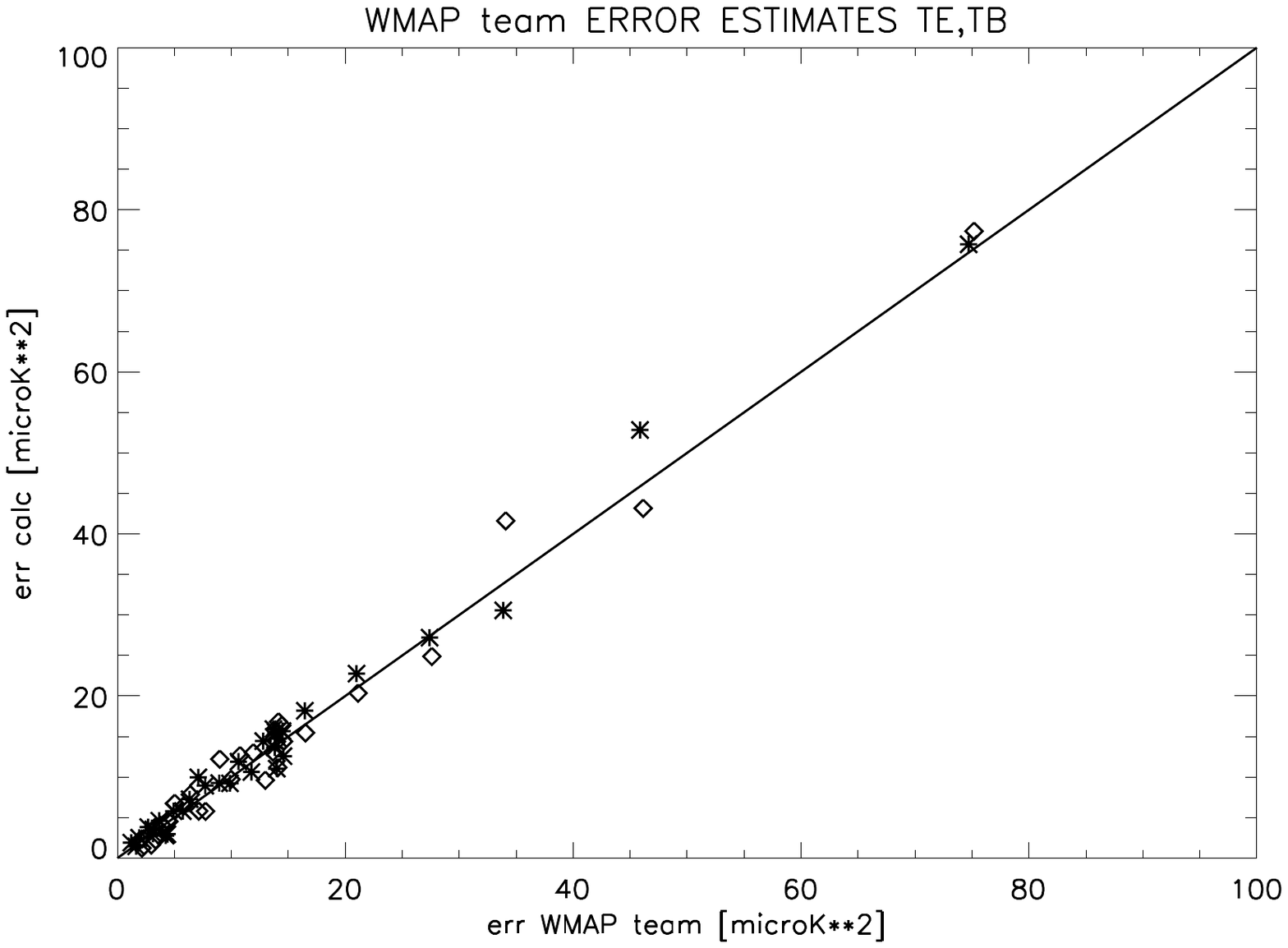}
\caption{The X-axis gives the errors of the power spectra derived by the WMAP Team from a detailed analysis of the noise distributions (incl. observational errors and cosmic variance). The Y-axis gives the errors determined from the scatter within each \emph{l} - interval (with more than 10 elements) of the individual power spectra derived by the WMAP Team. The one to one relation is shown. Symbols: TE (asterisks), TB (diamonds)}
\label{fig_sigma_team_te_tb}
\end{figure}

The WMAP team gives the TT, EE, BB , TE and TB power spectra for each multipole moment, as well as the binned power spectra with their estimated errors. In Figs.\ref{fig_sigma_team_tt_ee_bb} and \ref{fig_sigma_team_te_tb} the errors in the binned spectra, calculated from the scatter within each l-bin, are plotted  versus the estimate errors {(incl. observational and cosmic variance} given by the WMAP team. These estimates fit well. Therefore, since the emphasis is on the feasibility of the method and the rms estimates of the simple method are accurate enough for this purpose, the errors in the power spectra derived by the neural network method will be estimate by this method.

\section{Modeling the combined foreground spectrum}

A fundamental limitation in all CMB data analysis is the small number of frequency bands observed. In order not to boost the uncertainties in the derived parameters, only a small number of independent parameters can be applied. This is, of course, especially true for the WMAP data set, which contains fluxes in only 5 frequency bands (5 temperature and 5 x 2 polarization fluxes per sky pixel).

\subsection{The model of the spectral behaviour of the intensity of the foregrounds}
In the analysis of the WMAP 5yr temperature maps, NN10 used a simple model based on the spectral slopes: Ka/K, Q/Ka, V/Q, and W/V, where Ka/K is defined as
   \begin{equation}
   \rm
   Ka/K = log(flux(Ka)/flux(K)) / log(\nu(Ka)/\nu(K)).
   \end{equation}

   The other slopes are defined in a similar way.

To assure a reliable determination in NN10, the slopes was determined well within the Milky Way.
With the improved accuracy of the WMAP 7yr data and and by scaling the maps to nside = 128 and nside = 64, the following relations between these slopes has been found to be representative for fluxes from the bright part of the Milky Way to the areas covered be the KQ75 + Pol mask (defined in Section 6.1):
\begin{equation}
\rm
Q/Ka = -0.084 + 0.958 * Ka/K
\end{equation}
\begin{equation}
\rm
V/Q  ~~= +0.247 + 0.357 * Q/Ka
\end{equation}
\begin{equation}
\rm
W/V  ~~= +1.506 - 0.607 * K/Ka\\
\end{equation}

Small changes relative to the relations used in NN10 are seen. Eqs. 2 - 4 implies that the foreground spectrum can be calculated from 2 basic parameters: the flux in the K band and the slope Ka/K. Figs. 1-3 in NN10 demonstrate that these relations are well determined. The main problem is, of course, the scatter around the relations. This scatter is due to both observational errors and intrinsic scatter. In Section 5.1 a test is discussed, where accidental errors, corresponding to the observed scatter, are added to the relations.

\subsection{Models of the polarization of the foregrounds}
The WMAP team has used several models of the galactic foreground in their data reduction and analysis (Gold et al. 2009, Gold et al. 2011). In their analysis they fit simultaneously the temperature and polarization data, while the NN method only uses the polarization data. They assume that only the synchrotron and the dust emission is polarized. Briefly, they have used the following models (labels used in Figs. 21 - 24):

1. the "Team 2 comp" model assumes that the spectra of these components follow simple power laws, with slopes independent of frequency but allowed to vary spatially

2. the "Team spin" model (introducing a spinning dust model in the temperature fit, but no polarization) assumes fixed slopes for the synchrotron and dust emission

3. the "Team steep" model assumes logarithmic variation of the synchrotron slope with frequency

4. the "Team foregr removal" model assumes a fixed variation of the synchrotron and dust slopes as function of the frequency (Gold et al. 2009, Table 3). The WMAP Team uses this model to remove the foreground polarized signals from the observed Q and U maps before the CMB polarized signals are extracted.

Neural networks have been set up adopting the detailed assumptions of each of these models. An additional model (called "NN-temp") assumes the spectral behavior as described for the temperature model in Sect. 3.1.

As by the WMAP Team, the polarization direction and amplitude relative to the temperature flux are assumed to be independent of frequency, for all foreground models . The ranges of the parameters have been taken from the MCMC maps given by the WMAP Team.
 A flux unit of $10^{-20} erg/cm^{2}/s/Hz/sr$ has been used throughout this paper.

\section{Brief description of  the neural network concept}
\begin{figure}[!tbp]
\centering
\includegraphics[width=3.0 in]{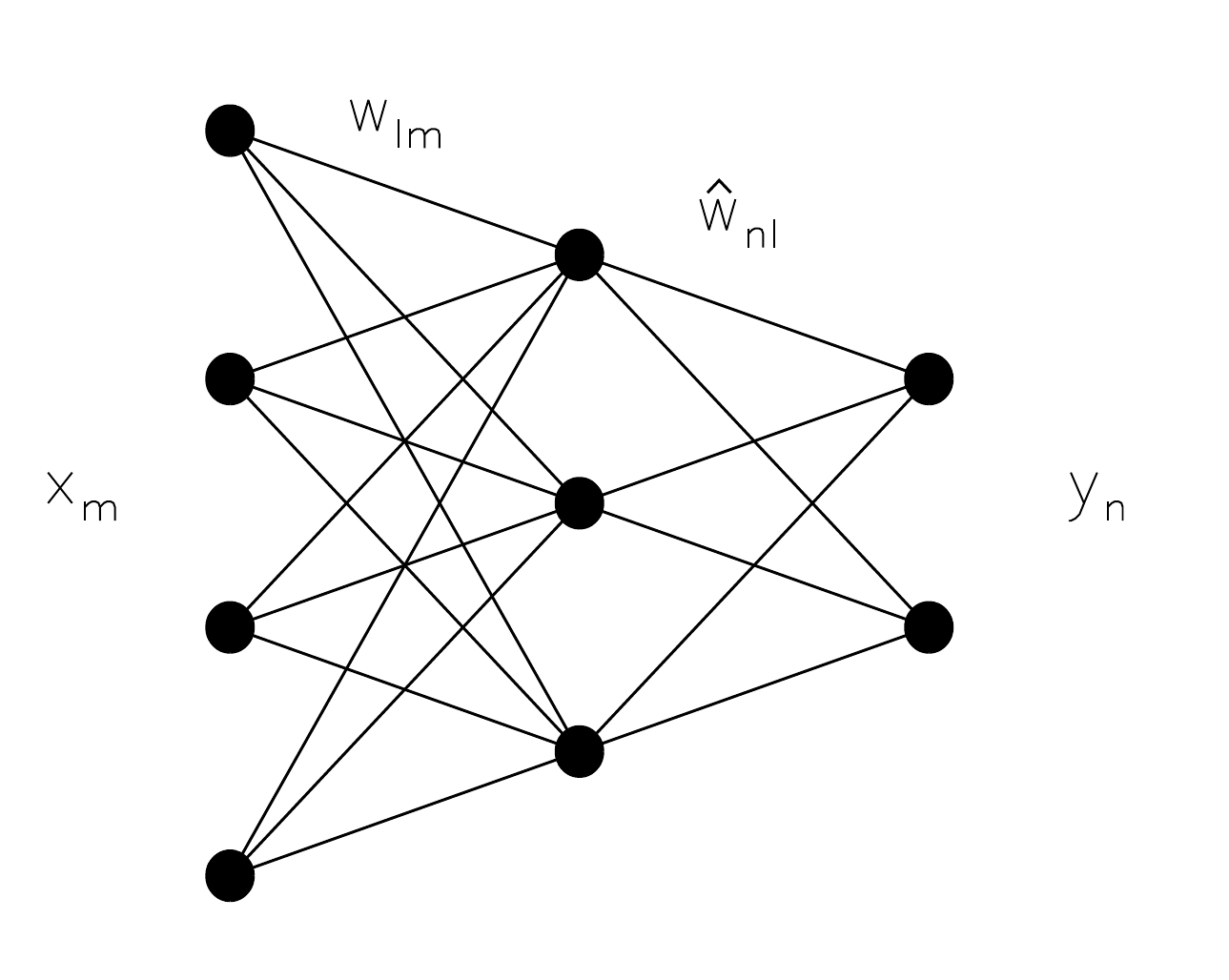}
\caption{A schematic of a multi layer perceptron network with
4 input channels, one hidden layer and 2 output channels}
\label{fignnet}
\end{figure}

    Neural networks are analogue computational systems whose structure is inspired by studies of the human brain. An excellent introduction to the many different types of neural networks can be found in Bishop\,\cite{bish95}. In the current paper, as for the previous papers NNJ, NNH and NN10, one of the simplest and also most popular networks, the multilayer perceptron (MLP), has been applied. Only a brief description of the neural network method will be given here.

    An MLP consists of a network of units (called processing elements, neurons, or nodes), conceptually illustrated in Fig.\ref{fignnet}. Each unit is shown as a circle and the lines connecting them are known as weights or links. The network can be understood as an analytical mapping between a set of input variables $x_{m}~(m = 1,...,M)$  and a set of output variables $y_{n}~(n=1,...,N)$. The input variables are applied to the M input units on the left of the figure: M=4 and N=2 in the shown example. These variables are multiplied by a matrix of parameters $w_{lm}~(l = 1,...,L;~
    m=1,...,M)$ corresponding to the first layer of links. Here L is the number of units in the middle (hidden) layer: L=3 in the shown example. This results in a vector of inputs to the units in the hidden layer. Each component of this vector is then transformed by a non-linear function F, giving
    \begin{equation}
    z_{l}~=~F \left( \sum_{m=1}^{M}~w_{lm}x_{m}~+\Theta_{l} \right) ~~(l=1,...,L), \label{eq1}
    \end{equation}
    where $\Theta_{l}$ is an offset or threshold. For the non-linear function F, the $tansig$ function has been chosen:
    \begin{equation}
    \mathrm{tansig}(x) ~= ~ \frac{2}{1~+~\mathrm{exp}(-2~x)}~-1. \label{eq2}
    \end{equation}
    It is seen that \emph{tansig} is highly non - linear, with values falling within the interval $[-1:1]$.
    From the hidden layer to the output units a linear transformation with weights $\widehat {w}_{nl}~(n=1,...,N;l=1,...,L)$ and offsets $\widehat{\Theta}_{n}$ are applied
    \begin{equation}
    y_{n}~=~\sum_{l=1}^{L}\widehat{w}_{nl}z_{l}~+~\widehat{\Theta}_{n}
		\quad\quad (n=1,...,N). \label{3}
    \end{equation}
    Combining Eqs.\,1 and 2 shows that the entire network transforms the inputs $x_{m}$ to the outputs $y_{n}$ by the following analytical function
    \begin{equation}
    y_{n}(x_{1},...,x_{M})~=~ \sum_{l=1}^{L}\widehat{w}_{nl}~F \left( \sum_{m=1}^{M}w_{lm}x_{m}~+~\Theta_{l}\right)~+~\widehat{\Theta}_{n}. \label{eq4}
    \end{equation}
    where F is the \emph{tansig} function (called activation function). Clearly, such an MLP can be easily generalized to more than one hidden layer.

    Given a set of P example input and output vector pairs $\{x_{m}^{p}~y_{n}^{p}\}~ p=1,...,P$ for a specific mapping, a technique known as error back propagation, can derive estimates of the parameters $w_{lm},~\Theta_{m}$ and   $\widehat{w}_{nl},~\widehat{\Theta}_{n}$, so that the network function (\ref{eq4}) will approximate the required mapping.
    The training algorithm minimizes the error function
    \begin{equation}
    E_{NN} ~=~\sum_{p=1}^{P}\sum_{n=1}^{N}[y_{n}(x^{p}) ~- ~y_{n}^{p}]^{2}. \label{eq5}
    \end{equation}

    A neural network is set up to handle a given data set. Traditionally, this is split into 3 data sets: one used directly to train the network, and a validation data set used in the iteration scheme, not directly in the training, but in the evaluation of the improvement of the network. And a third independent data set used only at the end of the training to get an estimate of the accuracy of the derived network.

\section{The applied neural network}
A basic assumption for this method is that the noise is white (i.e. no 1/f noise). If this is not the case, it is necessary to correct the maps for non - white features, before the data is run  through the network. With the assumption of white noise, the noise of the individual sky pixels is independent, and it is possible to treat each pixel separately.

The neural networks applied here have 10 input channels, two for each of the WMAP 5 frequencies and two output channels, the CMB Q and U. Together the 10 input values are referred to as a spectrum. The setup of the neural network follows this scheme:
\begin{enumerate}
\item To simulate a spectrum, draw relevant number of independent parameters, uniformly distributed, within specified ranges:

\item Calculate the resulting Q and U for the 5 WMAP bands from the foreground model in Sect.3

\item For each frequency, add random Gaussian noise calculated from the WMAP 7 yr hit maps and the error per hit given in the WMAP web-site.

\item Repeat 1--3 until the desired number of spectra ($N_{NNET}$) has been obtained. This data set is split into a set used directly to train the network and a set used for validation of the iteration scheme.
\item Train the neural network to find the transformation between the input spectra and the true CMB Q and U (known for each spectrum of the training data set).

\item Obtain an independent test sample of spectra by repeating 1--3 $N_{TEST}$ times

\item Run the $N_{TEST}$ spectra through the network to get an independent estimate of the reliability of the network, derived from the skewness and kurtosis of the distributions of residuals and the correlations of the residuals with the input parameters

\item If the derived network is working satisfactorily, meaning that the systematic errors and correlations on the independent test sample are as small as found in our previous investigations, run the WMAP 7yr data through the network.

\end{enumerate}

 An MLP with 2 hidden layers (5 and 3 processing elements, respectively, referred to as the NN network) was used for the data sets considered here.
The experience is that about 10,000 spectra are enough for the data set used to train the network.

\subsection{General tests of the neural networks}
In the previous papers it has been established that for temperature data the NN method extracts the CMB signal without pollution from the foreground emission. In Table \ref{tab:table1} it is seen that for an independent test data set the distributions of the residuals from the polarization neural network are indistinguishable from a Gaussian, and that the residuals are uncorrelated with the input parameters . Similar networks were set up to fit the Q and U parameters of the foreground model (referred to as the synchrotron and dust networks). Quite similar results, as presented in Table \ref{tab:table1} for the CMB network, were found.
Therefore, also for polarization data, the NN method give Gaussian error distributions and very small systematic errors in the extracted parameters

To investigate the sensitivity of the neural networks to deviations from the assumed spectral behaviour, a test sample, where random noise have been added to the 4 slopes of the NN-temp model (Section 3.1), has been run through the network derived from data where no noise has been added to the slopes. It turns out that adding random noise of 0.4 gives an insignificant increase the rms of the derived quantities, both for temperature and polarization samples.

\begin{table}
\caption{\label{tab:table1}
The statistics of the residuals of an independent test sample run through the CMB NN network.
The table contains the skewness and kurtosis of the residual distributions for the derived Q and U, together with the correlations of the residuals with the 5 input parameters}
%\begin{ruledtabular}
\begin{tabular}{crrrrrrc}
residuals & skew & kurt & $Q_{cmb}$& $U_{cmb}$& $Q_{for}$ & $U_{for}$& $Ka/K_{for}$\\

$\Delta$Q&  0.00 & +0.05 &  +0.09 & -0.02 & +0.06 & +0.01 & 0.00\\
$\Delta$U & 0.00 & +0.01 &  -0.02 & +0.07 & +0.00 & +0.04 & 0.00\\
\end{tabular}
%\end{ruledtabular}
\end{table}

\section{The results of the NN network}
\subsection{The polarization maps}

\begin{figure}[!tbp]
\centering
\includegraphics[angle=90, width=3.0 in]{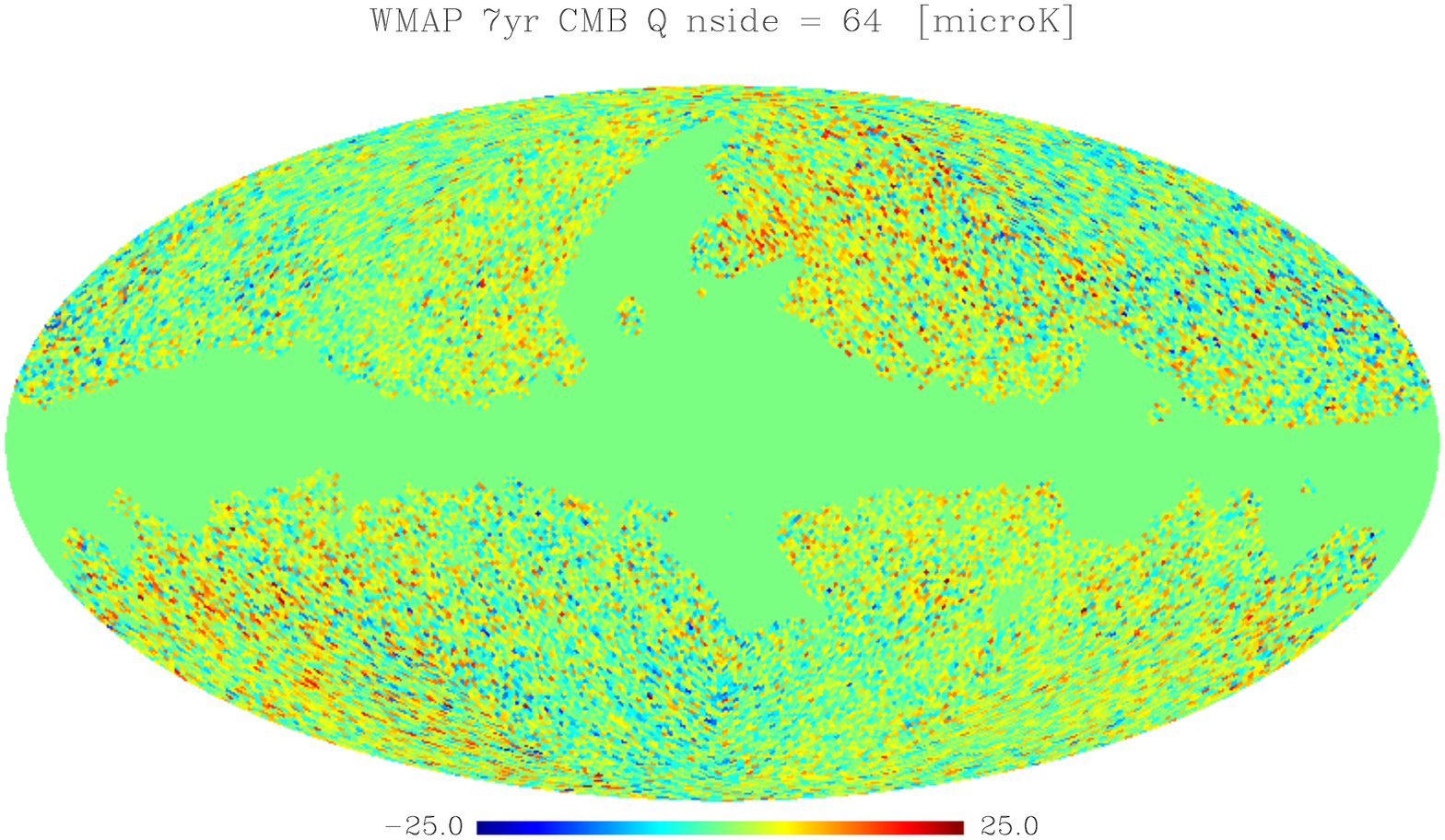}
\caption{The Q map derived by the CMB NN, color scale: $\pm$25 $\mu$K. The resolution is nside = 64. The KQ75 + Pol mask, applied in deriving the power spectra, is evident}
\label{fig_q_64}
\end{figure}

\begin{figure}[!tbp]
\centering
\includegraphics[angle=90, width=3.0 in]{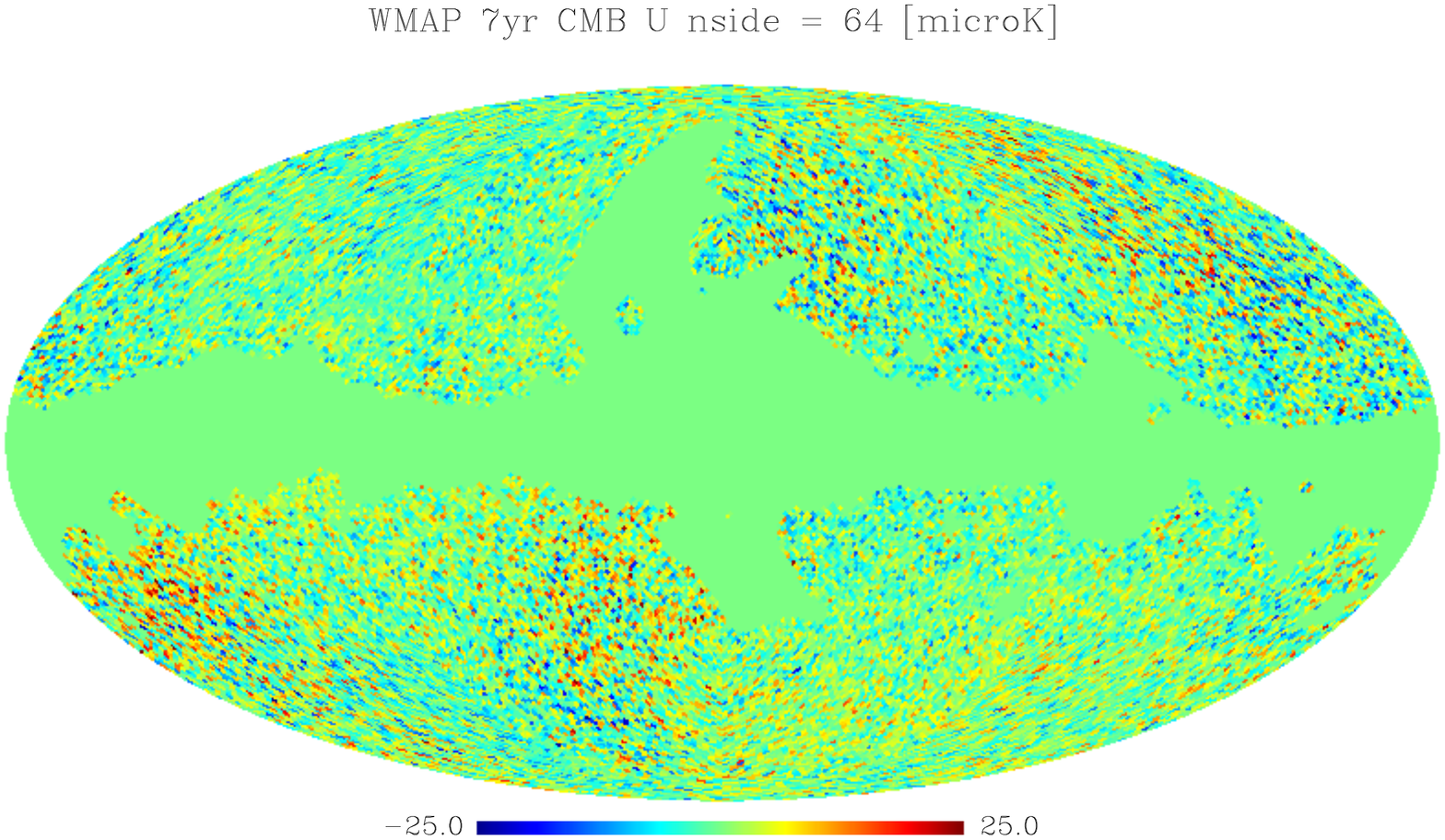}
\caption{The U map derived by the CMB NN, color scale: $\pm$25 $\mu$K. The resolution is nside = 64. The KQ75 + Pol mask, applied in deriving the power spectra, is evident}
\label{fig_u_64}
\end{figure}

\begin{figure}[!tbp]
\centering
\includegraphics[angle=90, width=3.0 in]{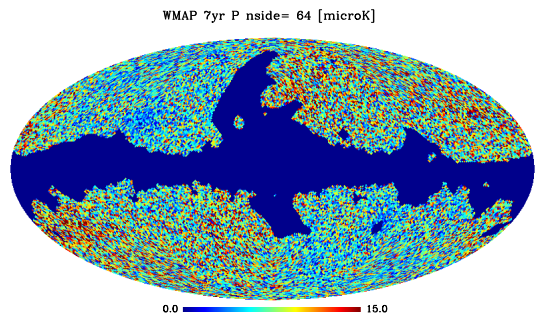}
\caption{The polarization amplitude map derived by the CMB NN, color scale: 0 - 15 $\mu$K. The resolution is nside = 64. The KQ75 + Pol mask, applied in deriving the power spectra, is evident. Also the areas of highest number of hits are apparent}
\label{fig_p_64}
\end{figure}
\begin{figure}[!tbp]
\centering
\includegraphics[angle=90, width=3.0 in]{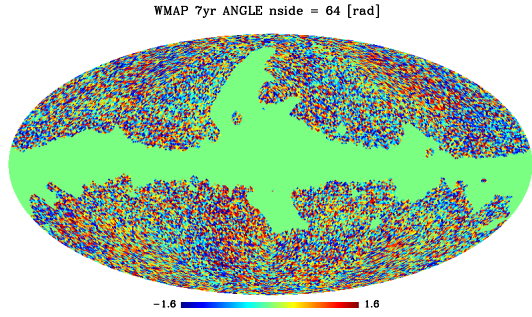}
\caption{The polarization direction map derived by the CMB NN, color scale: $\pm \pi$/2. The KQ75 + Pol mask, applied in deriving the power spectra, is evident.}
\label{fig_ang_64}
\end{figure}

\begin{figure}[!tbp]
\centering
\includegraphics[angle=90, width=3.0 in]{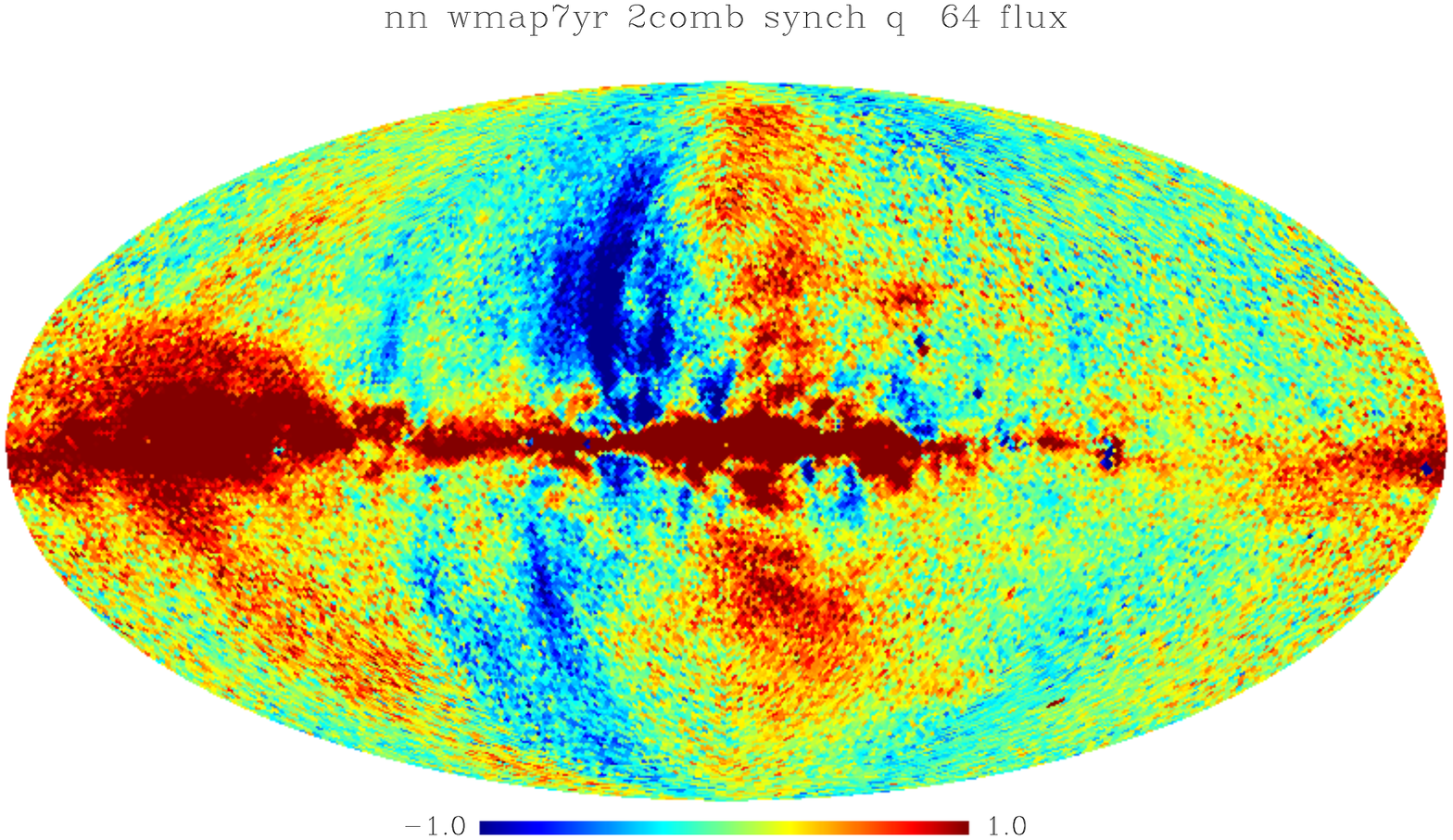}
\caption{The Q map in the K band derived by the foreground NN, nside = 64, range: $\pm$1 flux unit}
\label{fig_for_q_64}
\end{figure}

\begin{figure}[!tbp]
\centering
\includegraphics[angle=90, width=3.0 in]{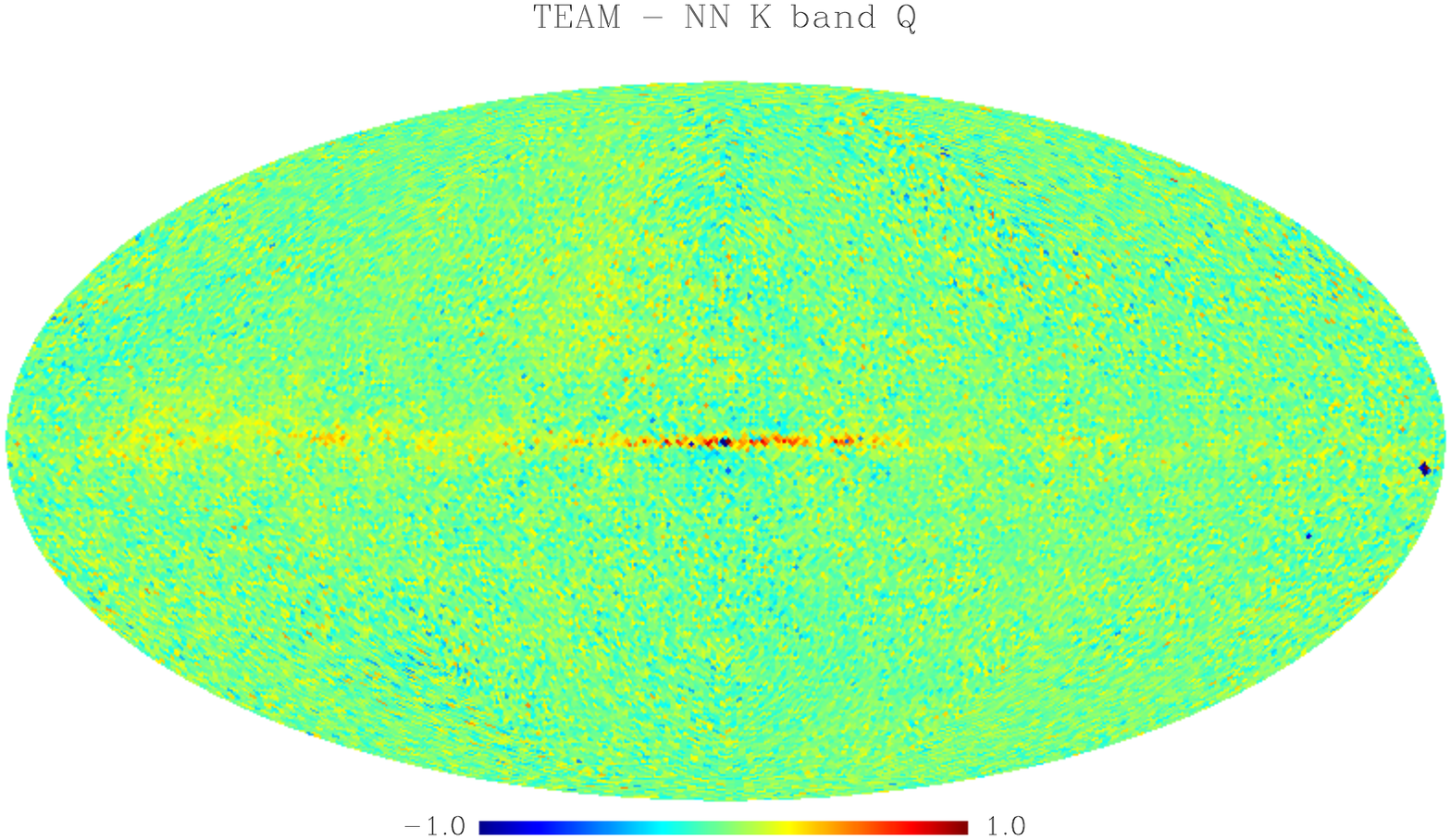}
\caption{The difference (WMAP Team - NN) foreground Q map in the K band, nside = 64, range: $\pm$1 flux unit}
\label{fig_res_for_q_64}
\end{figure}

\begin{figure}[!tbp]
\centering
\includegraphics[angle=90, width=3.0 in]{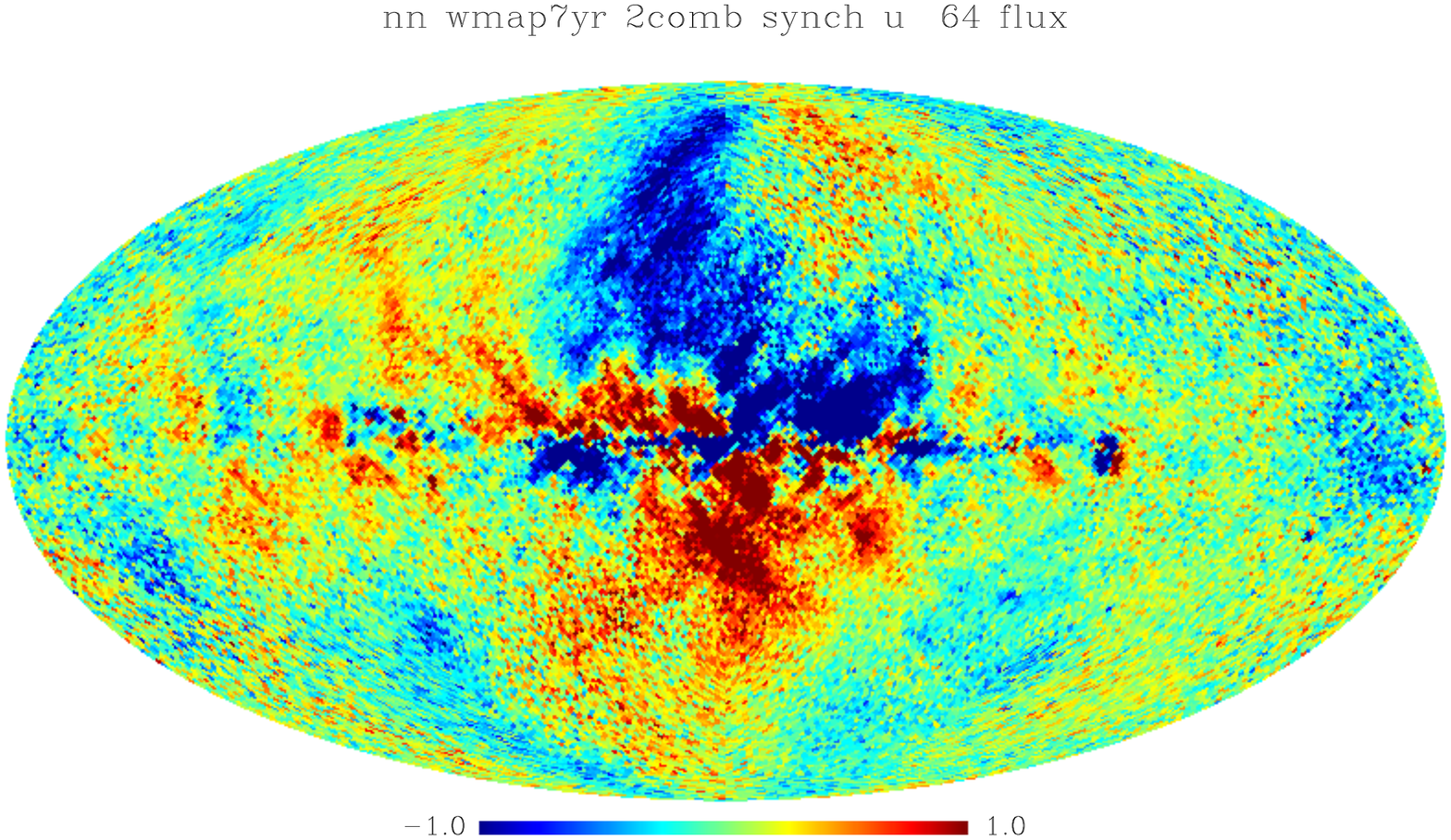}
\caption{The U map in the K band derived by the foreground NN, nside = 64, range:$\pm$1 flux unit}
\label{fig_for_u_64}
\end{figure}

\begin{figure}[!tbp]
\centering
\includegraphics[angle=90, width=3.0 in]{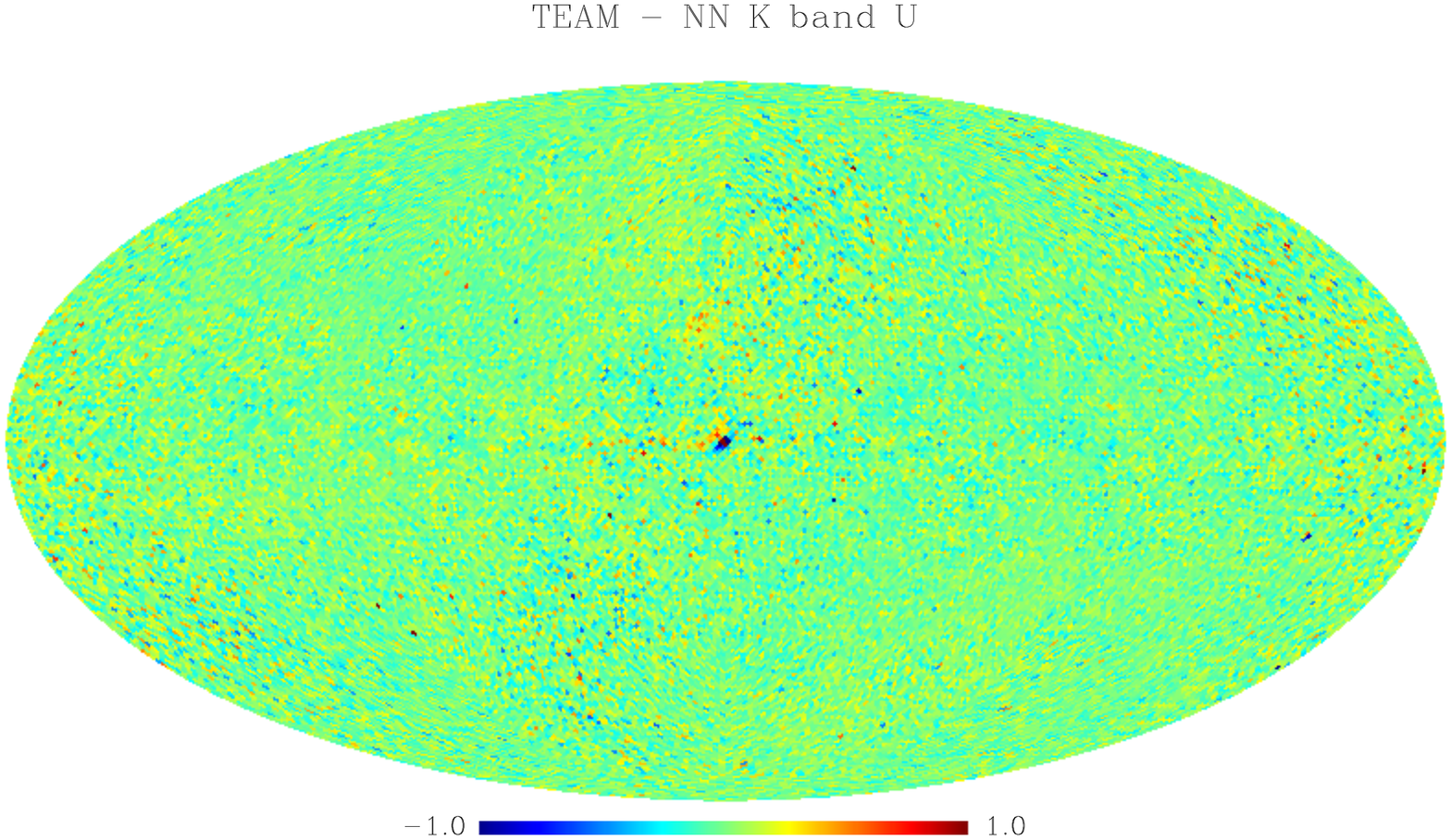}
\caption{The difference (WMAP Team - NN) foreground  U map in the K band, nside = 64, range:$\pm$1 flux unit}
\label{fig_res_for_u_64}
\end{figure}

\begin{figure}[!tbp]
\centering
\includegraphics[angle=90, width=3.0 in]{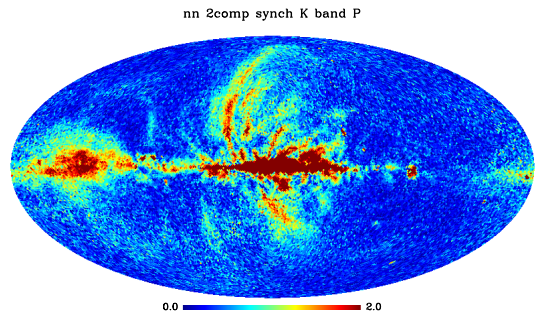}
\caption{The amplitude map in the K band derived by the foreground NN, nside = 64, range: 0 - 2 flux unit}
\label{fig_for_p_64}
\end{figure}

\begin{figure}[!tbp]
\centering
\includegraphics[angle=90, width=3.0 in]{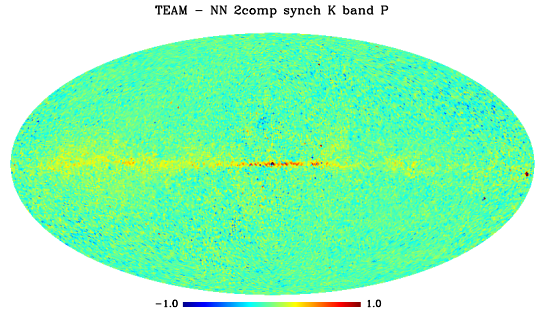}
\caption{The difference of the synch amplitude map in the K band determined by the WMAP Team and the foreground NN, range: $\pm$ 1.0 flux unit. }
\label{fig_res_p_64}
\end{figure}

\begin{figure}[!tbp]
\centering
\includegraphics[angle=90, width=3.0 in]{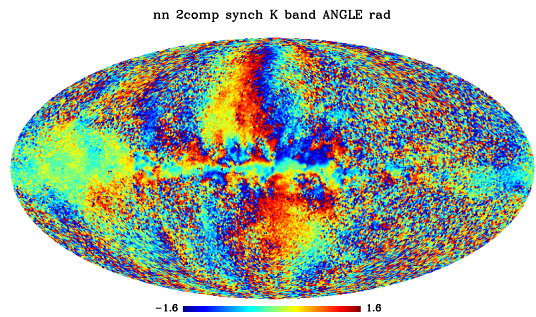}
\caption{The NN foreground direction map in the K band, nside = 64, range: $\pm \pi$/2}
\label{fig_for_ang_64}
\end{figure}

\begin{figure}[!tbp]
\centering
\includegraphics[angle=90, width=3.0 in]{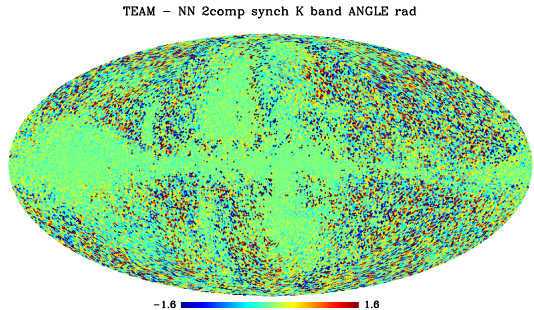}
\caption{The difference between polarization directions obtained by the WMAP team and the foreground NN, scale $\pm \pi$/2. }
\label{fig_res_ang_64}
\end{figure}

In the analysis of the WMAP 5yr temperature data (NN10) the TT power spectrum was derived within the WMAP Team KQ75 mask. Since this mask is not covering the North Polar Spur completely, this mask has been supplemented with the WMAP Team polarization mask (referred to as the KQ75 + Pol mask).

The WMAP masks have a sharp edge (only values 0 or 1). It is known that such masks can lead to mixing of E and B modes (Bunn et al.\cite{bunn03}; Lewis et al.  \cite{lewi02}). Following Kim\cite{kim010}, to minimize aliasing the combined mask has been convolved with a Gaussian with a 15 arcmin FWHM.

Figs. \ref{fig_q_64}, \ref{fig_u_64}, \ref{fig_p_64} and \ref{fig_ang_64} show the Q, U, polarization amplitude and direction maps derived from the signals extracted by the CMB NN. To improve statistical significance, the maps have been degraded to nside = 64. It is seen that there is no evidence for systematic errors in the maps. The KQ75 + Pol mask is evident in all figures.

Similarly, Figs. \ref{fig_for_q_64}, \ref{fig_for_u_64},\ref{fig_for_p_64} and \ref{fig_for_ang_64} show the same quantities derived with the synchrotron network. Figs. \ref{fig_res_for_q_64}, \ref{fig_res_for_u_64},\ref{fig_res_p_64} and  \ref{fig_res_ang_64} show the differences between the  MCMC maps obtained by the WMAP Team in the K band and the foreground NN. It is seen that the deviations are small, and that there is no strong correlation with the Q and U maps themselves, especially not within the KQ75 + Pol mask used to extract the power spectra in Section 6.2.
Similar results have been obtained with the dust network.

This indicates that the neural networks are able to disentangle the different \textbf{polarization} components (CMB, synchrotron, thermal dust) from each other, with no evidence for systematic errors.
\subsection{The NN polarization power spectra}

In this section we present the power spectra obtained with the NN network by means of the WMAP 7 yr data and compare them with the power spectra derived by the WMAP Team. By exploiting the HEALPix \emph{anafast} routine (Gorski et al. \cite{gors05}) the TT, EE, BB, TE, TB and EB power spectra have been extracted within the KQ75 + Pol mask.

The background noise spectra in the TT, EE and BB power spectra has been removed by exploiting the information given by the WMAP Team. The pixel noise is estimated from the number of hits for each individual sky pixel for the temperature maps, and from the noise covariance matrices for the Q and U maps. 30 sets of noise maps have been run through the NN networks. The average noise spectra have been adjusted with a small factor to fit the observed power spectra (for TT a factor of 1.14 was determined for 1150 $\leq$ l $\leq$ 1350, and for EE and BB a factor of 0.83 for 600 $\leq$ l $\leq$ 1200), implying a reasonable consistency of the WMAP noise model.

The corrections for the sky coverage of the KQ75 + Pol mask has been determined from 20 realization of noise power spectrum. The pixel size corrections have been taken from HEALPix. As demonstrated by Challinor et al. \cite{chal00}, by assuming pure co-polar beams, the polarized and unpolarized beams are the same, except for the very low l's. Since no information is available for the WMAP polarized beams, all power spectra have been corrected by the same window function. The effective window function used for correcting the 7yr power spectra have been determined in the same way as for the 5yr TT power spectrum (NN10).

\subsection{Tests for systematic errors in the derived power spectra}

\begin{figure}[!tbp]
\centering
\includegraphics[width=3.5 in]{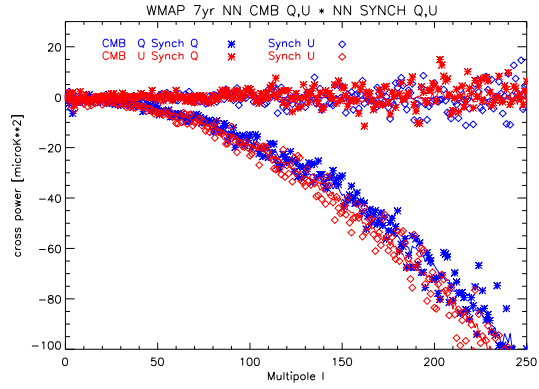}
\caption{The correlation power spectra between the Q and U derived by the CMB NN with the NN synchrotron Q and U. The power spectra around the zero - line is CMB U * Synch Q (red stars) and CMB Q * Synch U (blue diamonds) while the other spectra are CMB Q * Synch Q (blue stars) and CMB U * Synch U (red diamonds). It is seen that the spectra are following what is expected from pure correlated noise (the lines are the averaged simulated correlated noise)}
\label{fig_pow_nn_t_s_qu}
\end{figure}
\begin{figure}[!tbp]
\centering
\includegraphics[width=3.5 in]{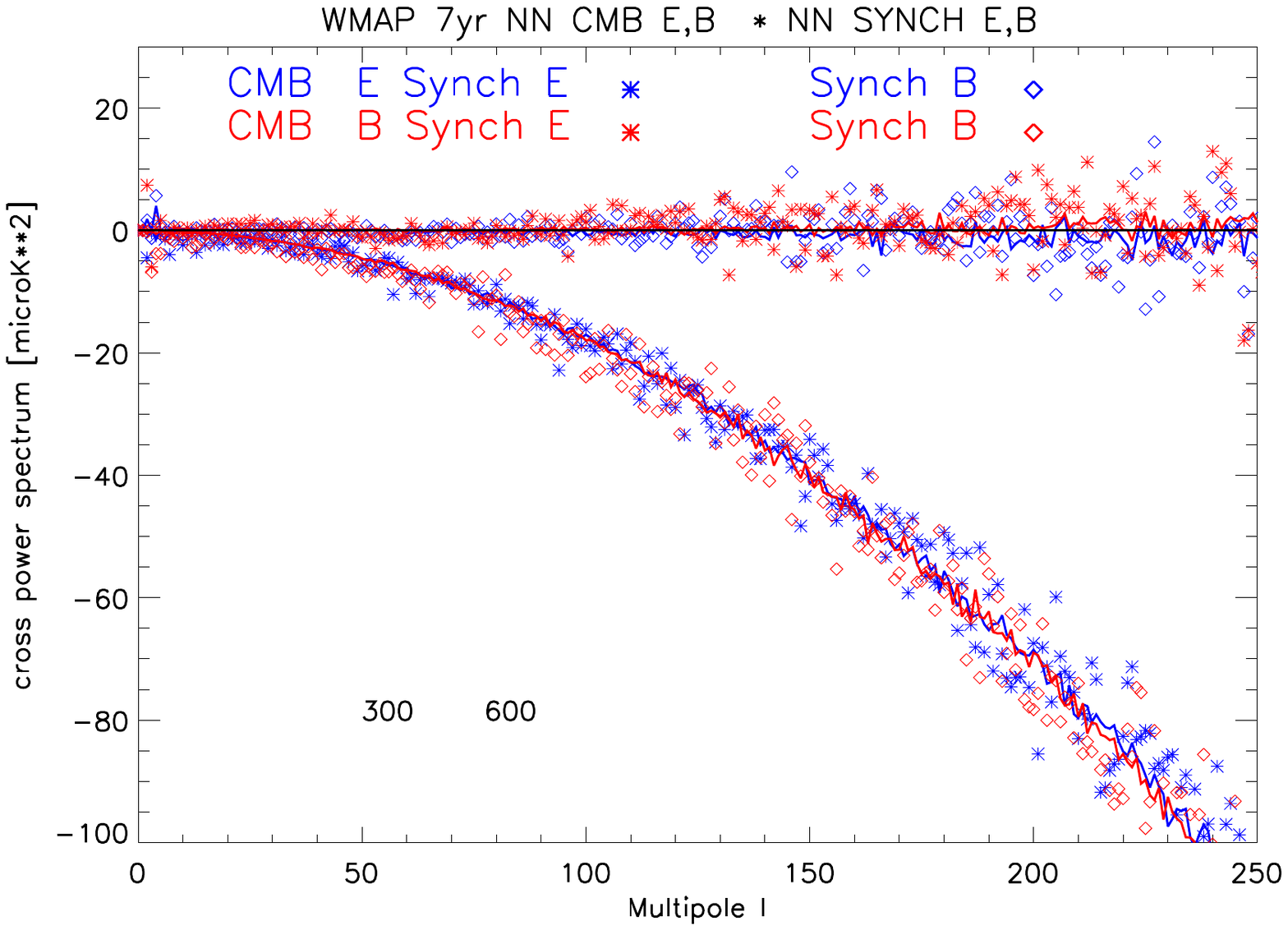}
\caption{The correlation power spectra between the E and B derived by the CMB NN with the NN synchrotron E and B. The power spectra around the zero - line is CMB B * Synch E (red stars) and CMB E * Synch B (blue diamonds) while the other spectra are CMB E * Synch E (blue stars) and CMB B * Synch B (red diamonds). It is seen that the spectra are following what is expected from pure correlated noise (the lines are the averaged simulated correlated noise)}
\label{fig_pow_nn_t_s_eb}
\end{figure}

\begin{figure}[!tbp]
\centering
\includegraphics[width=3.5 in]{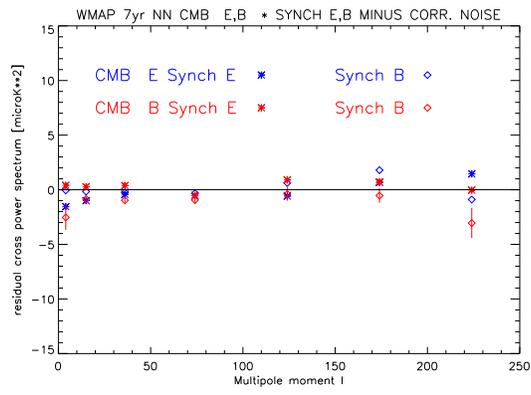}
\caption{The correlation power spectra between the E and B derived by the CMB NN with the NN synchrotron E and B with the correlated noise removed and averaged within the l ranges given by the WMAP team for the EE power spectrum. Same symbols as in Fig. 17 }
\label{fig_pow_nn_t_s_eb_res}
\end{figure}

\begin{figure}[!tbp]
\centering
\includegraphics[  width=3.0 in]{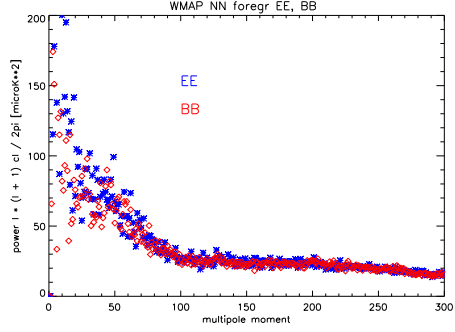}
\caption{The EE, BB power spectra of the foreground model in the K band, calculated from the PSM Challenge - 2 synchrotron map, applying the polarization amplitude and direction derived by the WMAP Team.}
\label{fig_foregr_ee_bb}
\end{figure}

\begin{figure}[!tbp]
\centering
\includegraphics[ angle = 90, width=3.0 in]{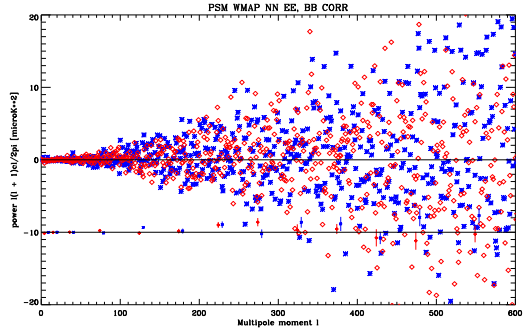}
\caption{The simulated, noise subtracted, EE (blue asterisks) BB (red diamonds) power spectra derived by the CMB NN, from maps calculated by means of the PSM reference CMB map, the PSM K synchrotron band map, the WMAP team polarization amplitude and direction maps. The frequency dependance of the NN-temp model in Section 3 has been used. The average power spectra incl errors are given, but shifted - 10 in Y for clarity}
\label{fig_psm_ee_bb_corr}
\end{figure}

\begin{figure}[!tbp]
\centering
\includegraphics[width=3.0 in]{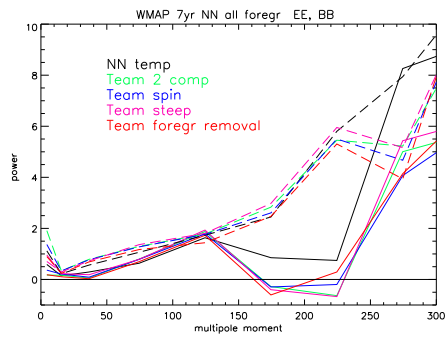}
\caption{The EE and BB power spectra derived by the CMB NN for all foreground models described in Section 3.2. The EE spectra are shown as solid line, while the BB spectra are given as dashed lines }
\label{fig_ee_bb_all_foregr}
\end{figure}

\begin{figure}[!tbp]
\centering
\includegraphics[width=3.0 in]{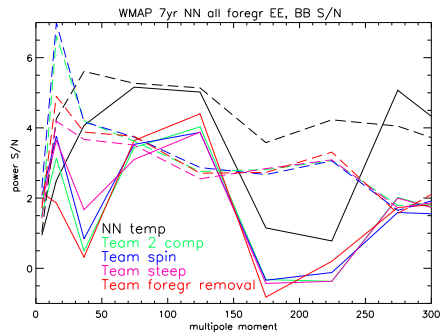}
\caption{The signal to noise ratios of the EE and BB power spectra (derived by the CMB NN) for all foreground models described in Section 3.2. The S/N of the EE spectra are shown as solid line, while the S/N of the BB spectra are given as dashed lines. It is seen that the NN-temp model provides better power spectra than the others}
\label{fig_ee_bb_all_foregr_sn}
\end{figure}

\begin{figure}[!tbp]
\centering
\includegraphics[width=3.0 in]{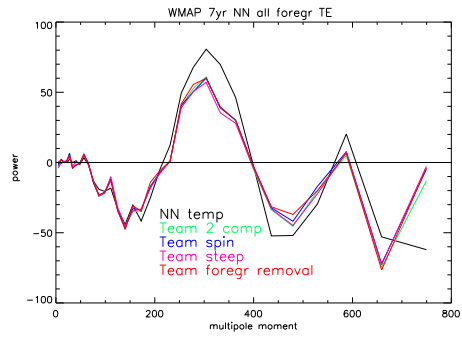}
\caption{The TE power spectra (derived by the CMB NN) for all foreground models described in Section 3.2. The TE spectra are shown as solid line}
\label{fig_te_all_foregr}
\end{figure}

\begin{figure}[!tbp]
\centering
\includegraphics[width=3.0 in]{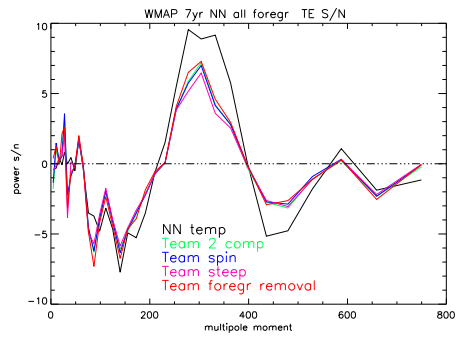}
\caption{The signal to noise ratios of the TE power spectra (derived by the CMB NN) for all foreground model described in Section 3.2. The S/N of the TE spectra are shown as solid line.It is seen that the NN-temp model provides better power spectra than the others }
\label{fig_te_all_foregr_sn}
\end{figure}

\begin{figure}[!tbp]
\centering
\includegraphics[ width=3.5 in]{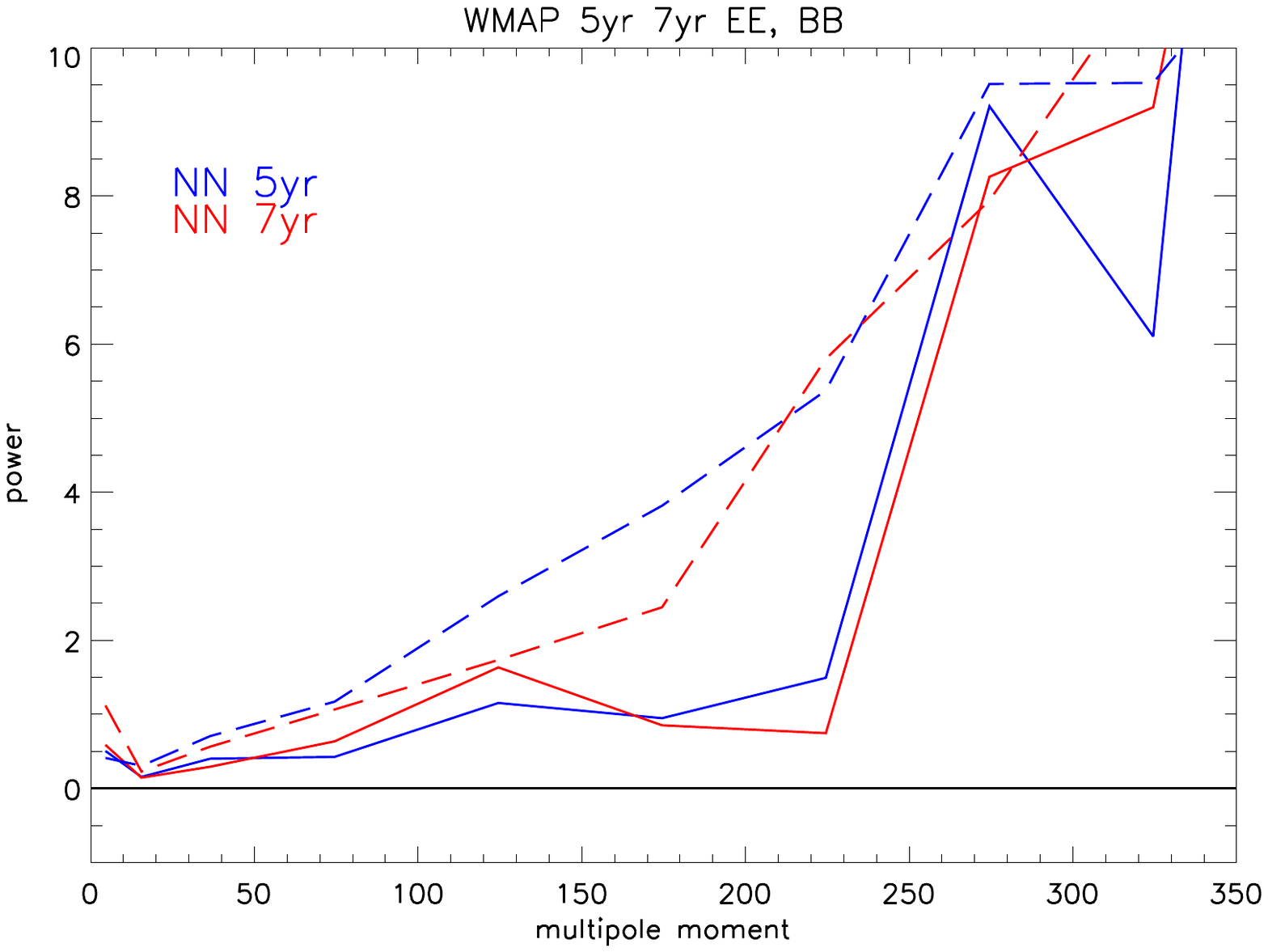}
\caption{The EE BB power spectra (derived by the CMB NN) for the WMAP 5yr and 7yr data. The EE spectra: solid lines. The BB spectra: dashed lines. 5yr spectra: blue lines. The 7yr spectra: red lines. Both data sets were run through the same CMB NN.}
\label{fig_5yr_7yr_ee_bb}
\end{figure}

Traditionally, the CMB maps have been analysed by means of power spectra, implying a lot of averaging of the data in the maps. Therefore, it is very sensitive to small non-Gaussianity features in the data and small systematic errors related to the foreground emission. In Section 5.1 it was demonstrated, as in the previous papers for temperature data, that the polarization networks are introducing neither non - Gaussian features nor systematic errors from the foregrounds in the extracted signals.

\subsubsection{Cross power spectra between the NN CMB Q and U maps and the NN foreground Q and U maps}

Of course, it is impossible to test the systematic errors in the derived NN maps for the real data, as it has been done for the simulated data. To investigate if significant systematic errors are present in the NN maps, the 4 cross power spectra between the Q,U derived by the CMB NN) and the NN synchrotron Q,U maps (within the KQ75 + Pol mask) have been investigated (Fig. \ref{fig_pow_nn_t_s_qu}). In order to estimate the expected noise cross power spectra, 20 realizations of  Q and U noise maps, derived from the noise properties given by the WMAP team, have been run through both the CMB and the synchrotron networks. The average cross power spectra are given in Fig. \ref{fig_pow_nn_t_s_qu} as straight lines. It is seen that the CMB NN and synchrotron NN cross power spectra (QQ,QU, UQ, UU) are completely dominated by correlated noise, excluding a significant pollution of the Q and U maps derived by the CMB NN. Fig \ref{fig_pow_nn_t_s_eb} show similar results for the EE, EB, BE and BB cross power spectra. In Fig. \ref{fig_pow_nn_t_s_eb_res} the EE and BB correlation noise have been removed, and the residual EE and BB power spectra, averaged within the l - ranges defined by the WMAP Team for the CMB EE power spectrum, are shown. It is seen that there is no sign of pollution of the E and B power spectra (CMB NN) from the synchrotron NN power spectra.
\subsubsection{Cross power spectra between the NN EE and BB spectra and simulated spectra}
To further investigate the possibility that the CMB NN network introduces systematic errors in the extracted E and B power spectra, simulations exploiting the maps prepared for the Planck WG2 Challenge-2 have been performed. The PSM noise - free synchrotron K band and the CMB reference maps have been used.
The polarization amplitude and direction found by the WMAP team for synchrotron emission has been assumed. Due to the large scatter in these maps, they have been smoothed with a 3 degree Gaussian. The foreground spectra for each sky pixel have been obtained using the "NN-temp" model explained in Section 3. As seen in Fig. \ref{fig_foregr_ee_bb} this model has strong E and B modes
The polarization amplitude and direction for the CMB have been assumed to be randomly distributed on the sky, which, of course, give no E and B modes.

The simulated maps have been convolved with the beam functions given by the WMAP Team, and realistic noise have been added. The resulting Q and U maps have been run through the CMB NN network. Fig. \ref{fig_psm_ee_bb_corr}, shows the extracted raw , noise subtracted EE and BB spectra. The average spectra including errors are also shown, displaced by - 10 in Y for clarity.  It is evident that the power spectra show no sign of pollution by the Galactic foregrounds.

\subsubsection{Power spectra obtained from the different models of the polarized foregrounds}
A basic problem for the NN methods is to assure that the assumed spectral behaviour of the foreground emission are covering the full data set to be analysed.  To investigate this problem, the WMAP 7yr data has been run through networks for each of the 5 foreground models described in Section 3.2. Figs. \ref{fig_ee_bb_all_foregr} and \ref{fig_ee_bb_all_foregr_sn} show for all the models the EE and BB power spectra and the S/N ratios derived by the CMB NN, respectively. Similarly, for the TE spectrum derived by the CMB NN) in Figs. \ref{fig_te_all_foregr} and \ref{fig_te_all_foregr_sn}. It is evident that these different foreground models give the same spectra, taking the errors into account. It is also evident that the 'NN-temp' model gives improved S/N power spectra compared to the other models.

The estimated S/N ratios of the extracted EE and BB power spectra are depending on the systematic errors in the mean noise spectra, subtracted from the total observed power spectra. For the l range [24,149] the S/N ratio of the EE power spectrum is 4.7, and 5.3 for the BB spectrum.  In order to reduce these S/N ratios to, say,  3.0, the level of the noise should be increased by 6 and  13 per cent, respectively. As stated above, the mean noise spectra have been adjusted to fit the observed spectra in the l range [600, 1200]. Such a large relative change in the scaling between the l range [24, 149] and the l range [600, 1200] is probably unreasonable, judged from the available information.

As a test of the overall consistency of the NN network, it is seen in Fig.\ref{fig_5yr_7yr_ee_bb} that the EE and BB  power spectra derived by the CMB NN extracted from the 5yr and 7yr data agree well with each other.

Altogether, it has been shown that the CMB neural networks are not introducing pollution from the foreground polarized emission in the extracted signals.
\subsection{The derived power spectra}

\begin{figure}[!tbp]
\centering
\includegraphics[ angle = 0, width=3.5 in]{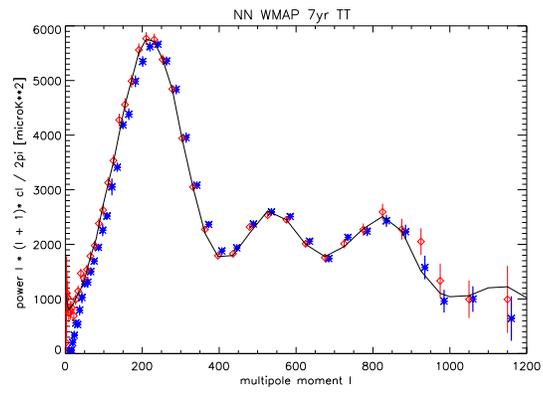}
\caption{The  TT 7yr power spectrum derived by the CMB NN. The blue asterisks with error bars give the NN power spectrum, while the red diamonds give the power spectrum of the WMAP Team. The solid black line  is the optimal $\lambda$CDM model found by the WMAP Team}
\label{fig_pow_tt}
\end{figure}

\begin{figure}[!tbp]
\centering
\includegraphics[angle=90,width=3.5 in]{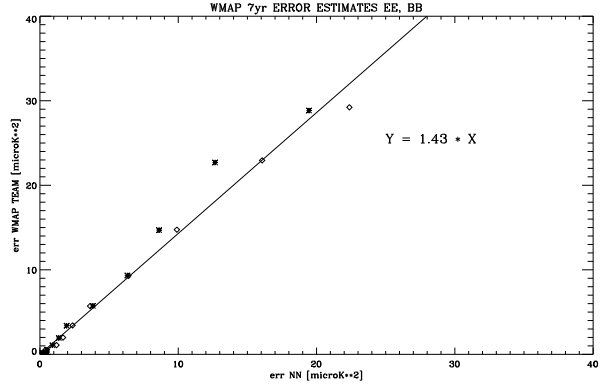}
\caption{The relation between the errors in the EE and BB power spectra derived by the CMB NN (x-axis) and the total errors (observational plus cosmic variance) by the WMAP Team (y-axis), for bins with more than 5 elements. It is seen that the NN errors are significantly smaller than the WMAP Team's errors}
\label{fig_sigma_nn_ee_bb}
\end{figure}

\begin{figure}[!tbp]
\centering
\includegraphics[ angle = 90, width=3.5 in]{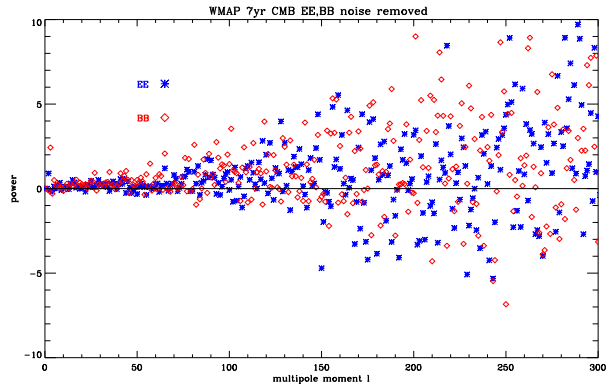}
\caption{The EE and BB power spectra derived by the CMB NN with the noise removed. The EE spectrum: blue asterisks, the BB spectrum: red squares It is seen that the distributions are significantly biased towards positive value, for EE up to $\sim$ 100 and BB $\sim$ 200}
\label{fig_wmap7yr_res_ee_bb}
\end{figure}

\begin{figure}[!tbp]
\centering
\includegraphics[ angle = 90, width=3.5 in]{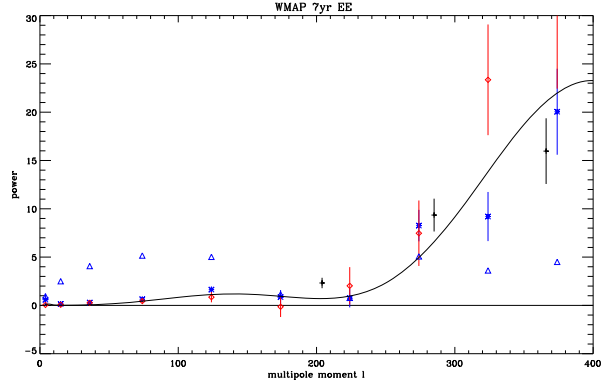}
\caption{The 7yr EE power spectrum derived by the CMB NN (blue asterisks with error bars). The S/N ratio of the spectrum is plotted as blue triangles (using just the numbers on the Y - scale). The WMAP EE 7yr power spectrum (Nolta et al.\cite{Nolt09}) is plotted as red diamonds with error bars. The black crosses with error bars are the results from the QUAD experiment (Brown et al. \cite{brow09}). It is seen that all data agrees taking the accuracy of the spectra into account. The black curve is the optimal $\lambda$CDM model found by the WMAP Team}
\label{fig_pow_ee}
\end{figure}

\begin{figure}[!tbp]
\centering
\includegraphics[ angle = 90, width=3.5 in]{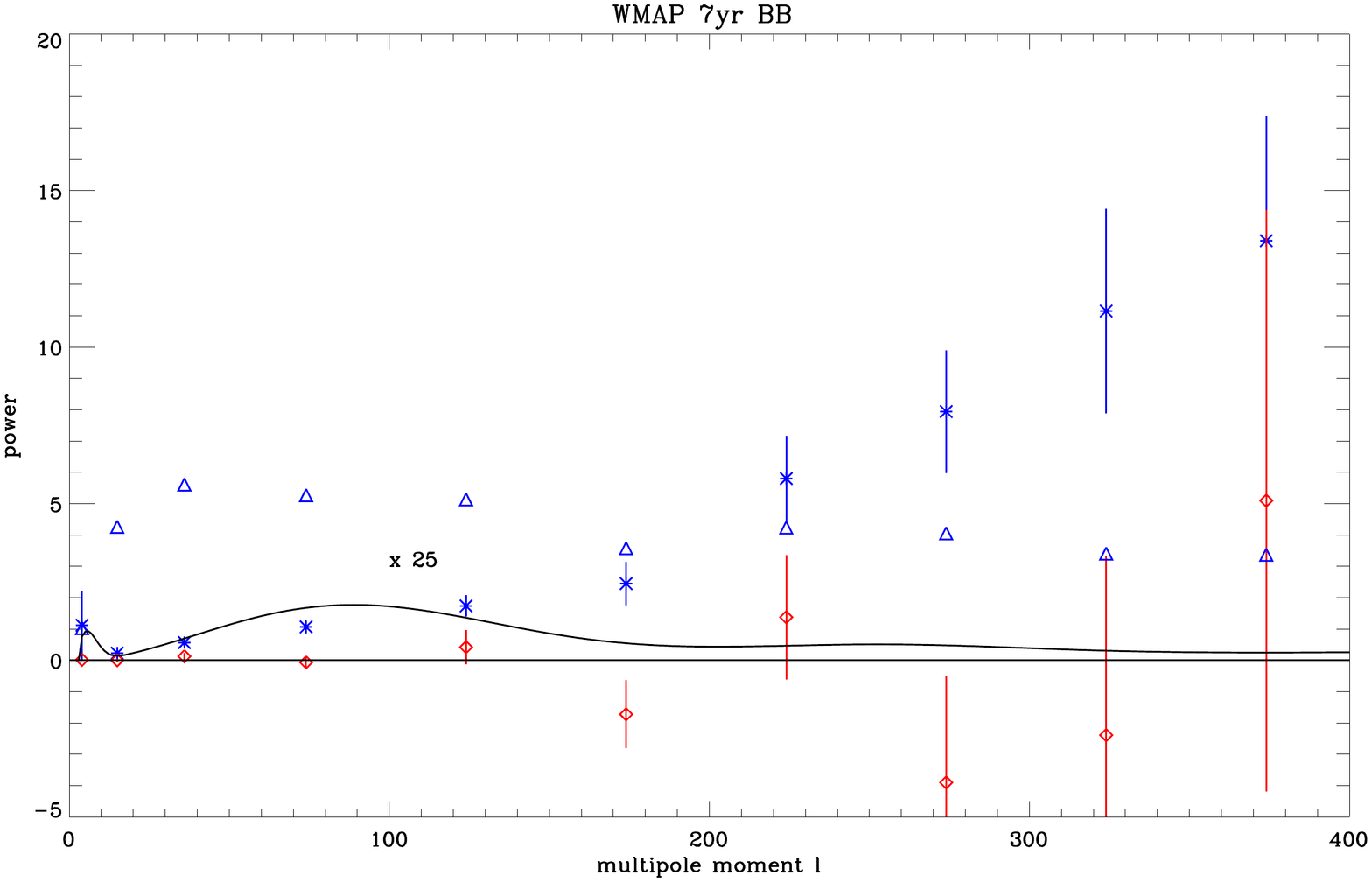}
\caption{The BB power spectrum derived by the CMB NN. The symbols are the same as in Fig.\ref{fig_pow_ee}.
The black curve is the BB spectrum derived from the optimal $\lambda$CDM model found by the WMAP Team, multiplied by a factor 25. It is seen that the NN spectrum is reliably detected out to l $\sim$ 300}
\label{fig_pow_bb}
\end{figure}

Fig.\ref{fig_pow_tt} shows the TT 7yr power spectrum derived by the CMB NN.
and the TT power spectrum of the WMAP Team. The NN errors are calculated as explained in Sect.II.B. It is seen that the NN TT power spectrum fit the theoretical model derived by the WMAP Team with good accuracy out to l=1200.

From Fig.\ref{fig_sigma_nn_ee_bb} it is seen that the errors in the EE and BB power spectra derived by the CMB NN are significantly smaller than the errors found by the WMAP team for the 7 yr data.

Fig.\ref{fig_wmap7yr_res_ee_bb} shows the raw, noise subtracted, EE and BB power spectra derived by the CMB NN. It is evident that the distributions are biased towards positive values up to l $\sim$100 and l $\sim$200 for EE and BB, respectively.

Fig.\ref{fig_pow_ee} shows the EE power spectrum derived by the CMB NN, together with the EE power spectrum obtained by the WMAP Team and  the best results from the QUAD experiment \cite{brow09}. For the NN EE spectrum, the signal-to-noise ratios are also given. It is seen that they fit, within the accuracy, with the optimal $\lambda$CDM model found by the WMAP Team. It is evident that the EE spectrum has been reliably detected by the CMB NN network up to l $\sim$150.

It is evident in Fig.\ref{fig_pow_bb} that the BB spectrum has also been detected with the CMB NN, up to l $\sim$ 300. For comparison, the prediction of the optimal $\lambda$CDM model by the WMAP Team multiplied by a factor of 25 is also shown (black line). Since the QUAD Team assumes that the BB spectrum is zero in their data reduction, their result are not included in Fig.\ref{fig_pow_bb}

\begin{figure}[!tbp]
\centering
\includegraphics[ width=3.5 in]{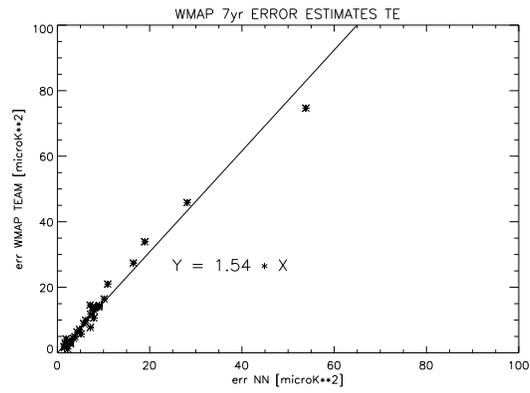}
\caption{The errors of TE (asterisks) and TB (diamonds) power spectra (derived by the CMB NN) (x-axis) compared
with the total errors (observational plus cosmic variance) derived by the WMAP Team (y-axis), for bins with more than 5 l's. It is seen that the NN errors are significantly smaller than the errors estimated by the WMAP Team }
\label{fig_sigma_nn_te_tb}
\end{figure}

Fig.\ref{fig_sigma_nn_te_tb} shows that also for the TE and TB power spectra (derived by the CMB NN) the errors are significantly smaller than the errors obtained by the WMAP Team.

\begin{figure}[!tbp]
\centering
\includegraphics[ angle=90, width=3.5 in]{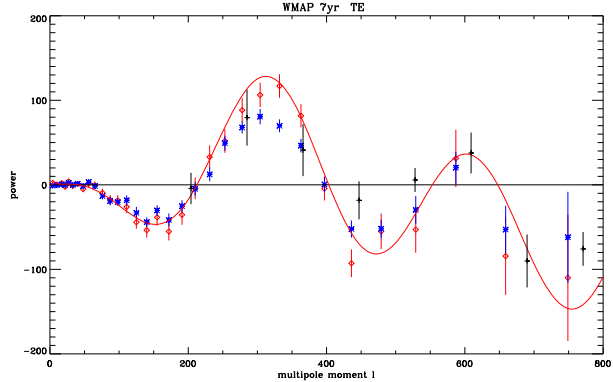}
\caption{The TE power spectrum derived by the CMB NN. The blue asterisks are the NN power spectrum, while the red diamonds are the power spectrum by the WMAP Team. The black crosses are the spectrum obtained by the QUAD Collaboration \cite{brow09}. }
\label{fig_pow_te}
\end{figure}

The TE power spectrum derived by the CMB NN) is shown in Fig. \ref{fig_pow_te} together with the spectra by the WMAP Team and the QUAD Team. It is seen that the NN spectrum fits reasonable well with the QUAD spectrum, while the amplitude at l $\sim$ 300 is significantly smaller than derived by the WMAP Team spectrum.

\begin{figure}[!tbp]
\centering
\includegraphics[ angle=90, width=3.5 in]{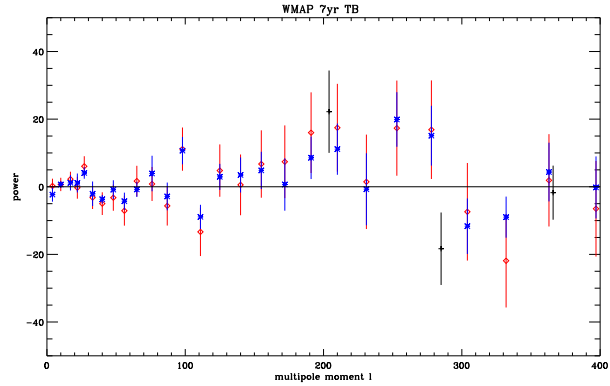}
\caption{The TB power spectrum (derived by the CMB NN). The blue asterisks with error bars are the NN power spectrum, while the WMAP Team spectrum are shown as red diamonds with error bars. The best results of the QUAD experiment are given as black crosses including error bars. It is evident that neither of the spectra represent an unambiguous detection}
\label{fig_pow_tb}
\end{figure}
\begin{figure}[!tbp]
\centering
\includegraphics[ angle=90,width=3.5 in]{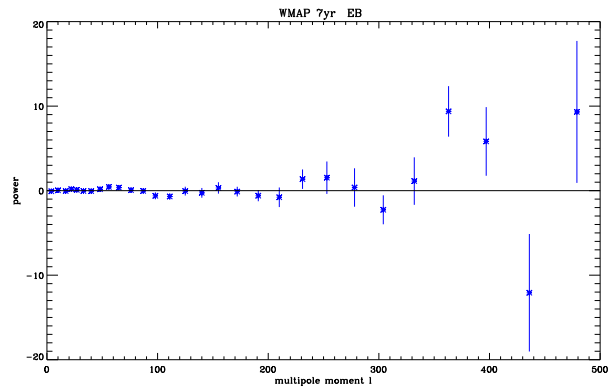}
\caption{The EB power spectrum (derived by the CMB NN, 0blue asterisks). It is evident that no reliable spectrum has been detected}
\label{fig_pow_eb}
\end{figure}

From Figs.\ref{fig_pow_tb} and \ref{fig_pow_eb} it is seen that neither the TB nor the EB power spectra (CMB NN) have been reliable detected.

\section{Conclusions}
By adopting the noise models and window functions defined by the WMAP Team and 5 different models of the polarized Galactic emission discussed by the WMAP Team, it has been demonstrated that the noise properties of the maps extracted by the NN are indistinguishable from Gaussian distributions, and that the extracted data is not significantly polluted by the foreground emission.

The TT power spectrum derived by the CMB NN fits well with the WMAP Team spectrum. The TE spectrum derived by the CMB NN fits the spectrum of the QUAD team well, but the power around l = 300 is smaller than found by the WMAP Team.

The errors in the power spectra of the signals extracted by the CMB NN network are significantly smaller than the errors obtained by the WMAP Team.

The CMB NN has detected both E and B modes, and they are not significantly polluted by the polarized signals from the Galactic foregrounds. The EE spectrum is consistent with the spectra found by the WMAP and QUAD teams, while the BB spectrum is significant stronger than found by the WMAP Team.
Although the BB spectrum has been extracted by the CMB NN and not polluted by the polarized foreground emission, a definitive prove of the origin must probably wait, until the analysis of the Planck polarization data has been finalized.

\acknowledgements
The author wants to thank Prof. G. Efstathiou for valuable discussions during the preparation of this paper. Drs. C. A. Oxborrow and A. Hornstrup are acknowledged for valuable comments to the paper.  The use of the Planck Sky Model, developed by subgroup of the Component Separation Working Group of the Planck Collaboration lead by J. Delabrouille, is acknowledged.

\end{document}